\documentclass[12pt]{article}

\usepackage{amsmath,amssymb,amsbsy,amsfonts,amsthm,latexsym,
                        amsopn,amstext,amsxtra,euscript,amscd,color}
\usepackage{color,xcolor}
\usepackage{latexsym}
\usepackage{cite}
\usepackage{amsfonts}
\usepackage{amsfonts,mathrsfs}
\usepackage{graphicx}
\usepackage{psfrag}
\usepackage{subfigure}
\usepackage{url}
\usepackage{stfloats}
\usepackage{amsmath}

\newtheorem{theorem}{Theorem}
\newtheorem{lemma}{Lemma}

\newcommand{\quash}[1]{}

\begin{document}

\def\F{\mathbb{F}}
\def\Z{\mathbb{Z}}

\title{On error linear complexity of new generalized cyclotomic binary sequences of period $p^2$}
\author{Chenhuang Wu$^{1,2}$, Chunxiang Xu$^{1}$, Zhixiong Chen$^{2}$, Pinhui Ke$^{3}$ \\
1. School of Computer Science and Engineering, \\
University of Electronic Science and Technology of China,\\
  Chengdu 611731, P. R. China\\
2. Provincial Key Laboratory of Applied Mathematics,\\
Putian University,\\
 Putian, Fujian 351100, P. R. China\\
3. Fujian Provincial Key Laboratory of Network Security and Cryptology, \\
College of Mathematics and Informatics,\\
 Fujian Normal University,\\
Fuzhou, Fujian, 350117, P. R. China}

%\date{}
\maketitle

\begin{abstract}
We consider the $k$-error linear complexity of a new binary sequence of period $p^2$, proposed in the recent paper
``New generalized cyclotomic binary sequences of period $p^2$", by Z. Xiao et al., who calculated the linear complexity of the sequences (Designs, Codes and Cryptography, 2017, https://doi.org/10.1007/s10623-017-0408-7).
More exactly, we determine the values of $k$-error linear complexity over $\mathbb{F}_2$ for almost $k>0$ in terms of the theory of Fermat quotients.
Results indicate that such sequences have good stability.
\end{abstract}



\section{Introduction}
Cyclotomic and generalized cyclotomic classes are widely adopted in cryptography. They play an important role in the design of pseudorandom sequences.
The typical examples are the Legendre sequences derived from cyclotomic classes modulo an odd prime
and the Jacobi sequences derived from generalized cyclotomic classes modulo a product of two odd distinct primes.
The generalized cyclotomic classes modulo a general number (such as a prime-power) are also paid attention in the literature, see the related  works such as \cite{CZHTY13,
DH98,DH99,W62,XCX16,ZCTY13} and references therein.

Recently, a new family of binary sequences were introduced by Xiao, Zeng, Li and Helleseth \cite{XZLH} via defining the generalized cyclotomic classes
modulo $p^2$ for odd prime $p$.
Now we introduce the way of defining generalized cyclotomic classes
modulo $p^2$.

Let $p-1=ef$ and $g$ be a primitive root\footnote{For our purpose, we will choose $g$ such that the fermat quotient $q_p(g)=1$, see the notion
in Sect. 2.} modulo $p^2$. The generalized cyclotomic classes for $1\le j\le 2$ is defined by
$$
D_0^{(p^j,f)}\triangleq \{g^{kfp^{j-1}} \pmod {p^j} : 0\leq k<e\}
$$
and
$$
D_l^{(p^j,f)}\triangleq g^{l} D_0^{(p^j,f)}=\{g^l\cdot g^{kfp^{j-1}} \pmod {p^j} : 0\leq k<e \}, ~~ 1\leq l<fp^{j-1}.
$$
Then The authors of \cite{XZLH} chose even $f$ and an integer $b\in \mathbb{Z} : 0\leq b<fp$ to define a new $p^2$-periodic binary sequence $(s_n)$:
\begin{equation}\label{def-binary}
s_n=\left\{
\begin{array}{ll}
0, & \mathrm{if} ~ n\pmod {p^2} \in \mathcal{C}_0,\\
1, &  \mathrm{if} ~  n\pmod {p^2} \in \mathcal{C}_1,
\end{array}
\right.
\end{equation}
where
$$
\mathcal{C}_0=\bigcup_{i=f/2}^{f-1}pD_{i+b \pmod f} ^{(p)}\cup\bigcup_{i=pf/2}^{pf-1} D_{i+b \pmod {pf}} ^{(p^2)}
$$
and
$$
\mathcal{C}_1=\bigcup_{i=0}^{f/2-1}pD_{i+b \pmod f} ^{(p,f)}\cup\bigcup_{i=0}^{pf/2-1} D_{i+b \pmod {pf}} ^{(p^2,f)}\cup \{0\}.
$$
The notation $pD_{j} ^{(p,f)}$ above means that $pD_{j} ^{(p,f)}=\{pv : v\in D_{j} ^{(p,f)}\}$.
They determined the linear complexity (see below for the notion) of the proposed sequences $(s_n)$ for $f=2^r$ for some integer $r\geq 1$.

\begin{theorem}\label{lc}(\cite[Thm. 1]{XZLH})
Let $(s_n)$ be the binary sequence of period $p^{2}$ defined in Eq.(\ref{def-binary}) with $f=2^{r}$ (integer $r>0$) and any $b$ for defining $\mathcal{C}_0$ and $\mathcal{C}_1$.
If $2^{(p-1)/f}\not \equiv 1 \pmod{p^2}$, then
the linear complexity  of $(s_n)$  is
\[
 LC^{\F_2}((s_n))=\left\{
\begin{array}{ll}
p^2-(p-1)/2, & \mathrm{if}\,\ 2\in D_0^{(p,f)}, \\
p^2, & \mathrm{if}\,\ 2\not\in D_0^{(p,f)}.
\end{array}
\right.\\
\]
\end{theorem}
The linear complexity is an important cryptographic characteristic of sequences
and provides information on the predictability and thus unsuitability for cryptography. Here we give a short introduction of the linear complexity of periodic sequences.
Let $\F$ be a field.  For a $T$-periodic
sequence $(s_n)$ over $\F$, we recall that the
\emph{linear complexity} over $\F$, denoted by  $LC^{\F}((s_n))$, is the least order $L$ of a linear
recurrence relation over $\mathbb{F}$
$$
s_{n+L} = c_{L-1}s_{n+L-1} +\ldots +c_1s_{n+1}+ c_0s_n\quad
\mathrm{for}\,\ n \geq 0,
$$
which is satisfied by $(s_n)$ and where $c_0\neq 0, c_1, \ldots,
c_{L-1}\in \mathbb{F}$.
Let
$$
S(X)=s_0+s_1X+s_2X^2+\ldots+s_{T-1}X^{T-1}\in \mathbb{F}[X],
$$
which is called the \emph{generating polynomial} of $(s_n)$. Then the linear
complexity over $\F$ of $(s_n)$ is computed by
\begin{equation}\label{licom}
  LC^{\F}((s_n)) =T-\deg\left(\mathrm{gcd}(X^T-1,
  ~S(X))\right),
\end{equation}
see, e.g. \cite{CDR} for details.

For a sequence to be cryptographically strong, its linear complexity
should be large, but not significantly reduced by changing a few
terms. This directs to the notion of the $k$-error linear complexity.
For integers $k\ge 0$, the \emph{$k$-error linear complexity} over $\F$ of $(s_n)$, denoted by $LC^{\F}_k((s_n))$, is the smallest linear complexity (over $\F$) that can be
obtained by changing at most $k$ terms of the sequence per period, see \cite{SM}, and see \cite{DXS} for the related even earlier defined sphere complexity.  Clearly $LC^{\F}_0((s_n))=LC^{\F}((s_n))$ and
$$
T\ge LC^{\F}_0((s_n))\ge LC^{\F}_1((s_n))\ge \ldots \ge LC^{\F}_w((s_n))=0
$$
when $w$ equals the number of nonzero terms of $(s_n)$ per period, i.e., the weight of $(s_n)$.\\

The main contribution of this work is to determine the $k$-error linear complexity of $(s_n)$ in Eq.(\ref{def-binary}) for any even number $f$ (including $f=2^r$ considered in \cite{XZLH}).
The main results are presented in the following two theorems. The proofs appear in Section \ref{proofmain}.
Some necessary lemmas are introduced in Section \ref{lemmas}. A crucial tool for the proof is the {\em Fermat quotients}, which is introduced in Section \ref{fermat}.

\begin{theorem}\label{klc-2-primitive}(Main theorem)
Let $(s_n)$ be the binary sequence of period $p^{2}$ defined in Eq.(\ref{def-binary}) with even $f$ and any $b$ for defining $\mathcal{C}_0$ and $\mathcal{C}_1$. If $2$ is a primitive root modulo $p^2$, then
the $k$-error linear complexity of $(s_n)$  satisfies
\[
 LC^{\F_2}_k((s_n))=\left\{
\begin{array}{cl}
p^2, & \mathrm{if}\,\ k=0, \\
p^2-1, & \mathrm{if}\,\ 1\le k<(p-1)/2, \\
p^2-p, & \mathrm{if}\,\ (p-1)/2\leq k<(p^2-p)/2,\\
p-1, & \mathrm{if}\,\ k=(p^2-p)/2,\\
1, & \mathrm{if}\,\ k=(p^2-1)/2,\\
0, & \mathrm{if}\,\ k\geq (p^2+1)/2,
\end{array}
\right.\\
\]
if $p\equiv 3 \pmod 8$, and
\[
 LC^{\F_2}_k((s_n))=\left\{
\begin{array}{cl}
p^2, & \mathrm{if}\,\ k=0, \\
p^2-1, & \mathrm{if}\,\ 1\le k<(p-1)/2, \\
p^2-p+1, & \mathrm{if}\,\ k=(p-1)/2, \\
p^2-p, & \mathrm{if}\,\  (p+1)/2\leq k <(p^2-p)/2, \\
p, & \mathrm{if}\,\ k=(p^2-p)/2,\\
1, & \mathrm{if}\,\ k=(p^2-1)/2,\\
0, & \mathrm{if}\,\ k\geq (p^2+1)/2,
\end{array}
\right.\\
\]
if $p\equiv 5 \pmod 8$.
\end{theorem}

\section{Fermat quotients}\label{fermat}

In this section, we interpret that the construction of $(s_n)$ in Eq.(\ref{def-binary})  is related to Fermat quotients.
Certain similar constructions can be found in \cite{Chen,CD,CG,CNW,COW,GW}.

For integers $u\geq 0$, the {\it
Fermat quotient\/ $q_p(u)$} is the value in $\{0,1,\ldots, p-1\}$ at $u$ defined by
$$
q_p(u) \equiv \frac{u^{p-1} -1}{p} \pmod p,
$$
where $\gcd(u,p)=1$, if $p|u$ we set $q_p(u) = 0$, see \cite{OS}.

Thanks to the facts that
\begin{equation}\label{qp}
 \left\{
\begin{array}{ll}
q_p(u+\ell p)\equiv q_p(u)-\ell u^{-1} \pmod p,\\
q_p(uv)\equiv q_p(u)+q_p(v) \pmod p,
\end{array}
\right.
\end{equation}
for $\gcd(u,p)=1$ and $\gcd(v,p)=1$, we define
$$
D_{l}=\{u : 0\le u<p^2, \gcd(u,p)=1, q_p(u)=l\}, ~~ 0\le l<p.
$$
 In fact, together with
the second equation in (\ref{qp}) and the primitive root $g$ modulo $p^2$ with $q_p(g)=1$\footnote{Such $g$  always exists.},
we have
$$
D_l=\{g^{l+i p} \pmod {p^2} : 0\le i <p-1\},   ~~ 0\leq l<p.
$$
So according to the definition of $D_{l}^{(p^2,f)}$ in Sect. 1, we see that
\begin{equation}\label{DUD}
D_l= \bigcup_{i=0}^{f-1} D_{l+ip}^{(p^2,f)},   ~~ 0\leq l<p.
\end{equation}

The $D_l$'s and Eq.(\ref{qp}) help us to study the $k$-error linear complexity of $(s_n)$ in Eq.(\ref{def-binary}) in this work.

\section{Auxiliary lemmas}\label{lemmas}

In this section, we present some necessary lemmas needed in the proofs.
In the sequel, the notation $|Z|$ denotes the cardinality of the set $Z$.

\begin{lemma}\label{Dl}
Let $D_{l}$ be defined for $0\leq l<p$ by Fermat quotients as in Sect.2.
Then we have for $0\le l<p$,
$$
\{n \bmod p : n\in D_{l}\}=\{1,2,\ldots,p-1\}.
$$
\end{lemma}
Proof. Since $D_l=\{g^{l+i p} \pmod {p^2} : 0\le i <p-1\}$ for $0\leq l<p$,
we get
$$
\{g^{l+i p} \pmod {p} : 0\le i <p-1\}=\{g^{l+i} \pmod {p} : 0\le i <p-1\},
$$
which completes the proof.\qed

\begin{lemma}\label{fermat-D}
Let $D_{l}$ be defined for $0\leq l<p$ by Fermat quotients as in Sect.2.
Let $v\in \{1,2,\ldots,p-1\}$ and $\mathcal{V}_v=\{v, v+p,v+2p, \ldots, v+(p-1)p\}$.
Then for each $0\le l<p$, we have  $|\mathcal{V}_v \cap D_{l}|=1$.
\end{lemma}
Proof.
By the first equation in Eq.(\ref{qp}), if $q_p(v+i_1 p)=q_p(v+i_2 p)=l$ for $0\leq i_1, i_2<p$, that is $q_p(v)-i_1v^{-1}\equiv q_p(v)-i_2v^{-1} \pmod p$, then $i_1=i_2$.  \qed

\begin{lemma}\label{DDmodp}
Let $D_{l} ^{(p^j,f)}$ be defined for $0\leq l<fp^{j-1}$ with even $f$ and $1\le j\le 2$ as in Sect.1.
Then we have for any $0\le l<fp$
$$
\{u \pmod p : u \in D_{l} ^{(p^2,f)}\}=D_{l\pmod f} ^{(p,f)}.
$$
\end{lemma}
Proof. It is clear. \qed\\

\begin{lemma}\label{fermat-D-add}
Let $D_{l} ^{(p^j,f)}$ be defined for $0\leq l<fp^{j-1}$ with even $f$ and $1\le j\le 2$ as in Sect.1.
Let $v\in \{1,2,\ldots,p-1\}$ and $\mathcal{V}_v=\{v, v+p,v+2p, \ldots, v+(p-1)p\}$.
 If $v \in D_{\ell} ^{(p,f)}$ for some $0\le \ell<f$, we have for any $0\le l<p$
$$
\left|\mathcal{V}_v \cap D_{l+ip} ^{(p^2,f)}\right|=\left\{
\begin{array}{cl}
1,& \mathrm{if}~~i\equiv l-\ell \pmod {f},\\
0,& \mathrm{otherwise},
\end{array}
\right.
$$
where $0\leq i<f$.
\end{lemma}
Proof. For each $0\leq l<p$, we have by Lemma \ref{fermat-D}  $|\mathcal{V}_v \cap D_{l}|=1$. Then by Eq.(\ref{DUD})
there exists some $i_0 : 0\leq i_0<f$ such that
$$
|\mathcal{V}_v \cap D_{l+i_0p} ^{(p^2,f)}|=1.
$$
Now we determine $i_0$. Let $u\in \mathcal{V}_v \cap D_{l+i_0p} ^{(p^2,f)}$. We check that
$$
u\equiv v \pmod p, ~~~ u \pmod p \in D_{l+i_0p\pmod f} ^{(p,f)}
$$
by Lemma \ref{DDmodp}, hence we get $\ell \equiv l+i_0p\equiv l+i_0\pmod f$, that is $i_0\equiv l-\ell \pmod {f}$. We complete the proof. \qed\\

For any integer $b$, let
\begin{equation}\label{QN}
\mathcal{Q}_b=\bigcup_{i=0}^{f/2-1} D_{i+b \pmod f} ^{(p,f)} \subseteq \{1,2,\ldots,p-1\}, ~~~\mathcal{N}_b= \{1,2,\ldots,p-1\}\setminus \mathcal{Q}_b.
\end{equation}
We see that $|\mathcal{Q}_b|=|\mathcal{N}_b|=(p-1)/2$.

\begin{lemma}\label{card-CC-add}
Let $D_{l} ^{(p^j,f)}$ be defined for $0\leq l<fp^{j-1}$ with even $f$ and $1\le j\le 2$, and let $\mathcal{C}_0$ and $\mathcal{C}_1$ be defined with any integer $b$ as in Sect.1.
Let $v\in \{1,2,\ldots,p-1\}$ and $\mathcal{V}_v=\{v, v+p,v+2p, \ldots, v+(p-1)p\}$.

(1). If $v \in  \mathcal{Q}_b$, which is defined in Eq.(\ref{QN}), we have
$$
|\mathcal{V}_v \cap \mathcal{C}_0|=(p-1)/2, ~~ |\mathcal{V}_v \cap \mathcal{C}_1|=(p+1)/2.
$$

(2). If $v\in \mathcal{N}_b$,  which is defined in Eq.(\ref{QN}), we have
$$
|\mathcal{V}_v \cap \mathcal{C}_0|=(p+1)/2, ~~ |\mathcal{V}_v \cap \mathcal{C}_1|=(p-1)/2.
$$
\end{lemma}
Proof. It follows from Lemma \ref{fermat-D-add}. \qed\\

Define polynomials in $\mathbb{F}_2[X]$
\begin{equation}\label{d-poly}
 d_{l}^{(p^j,f)}(X)=\sum\limits_{n\in D_{l}^{(p^j,f)}}X^{n}
\end{equation}
 for $0\leq l<fp^{j-1}$ and $1\leq j\leq 2$. Let $\theta\in \overline{\mathbb{F}}_2$ be a primitive $p$-th root of unity
 and
$$
\omega_b=\sum\limits_{i=0}^{f/2-1}d_{i+b\pmod f}^{(p,f)}(\theta)=\sum\limits_{n\in \mathcal{Q}_b}\theta^n \in \overline{\mathbb{F}}_2.
$$
It is easy to see that 
$$
\omega_{b+f/2}=\sum\limits_{n\in \mathcal{N}_b}\theta^n ~~~~ \mathrm{and} ~~~~ \omega_b+\omega_{b+f/2}=1.
$$

\begin{lemma}\label{www-add}
Suppose that 2 is a primitive root modulo $p$.  Let $\theta\in \overline{\mathbb{F}}_2$ be a primitive $p$-th root of unity and 
$\omega_b=\sum\limits_{n\in \mathcal{Q}_b}\theta^n$. Then
$\omega_b\not\in \mathbb{F}_2$ for any integer $b$. 
\end{lemma}
Proof. 
Write
$$
W^{(1)}_b(X)=\sum\limits_{n\in \mathcal{Q}_b}X^n, ~~~ W^{(2)}_b(X)=\sum\limits_{n\in \mathcal{N}_b}X^n.
$$
Then we have $1<\deg(W^{(1)}_b(X)) \neq \deg(W^{(2)}_b(X))\leq p-1$.

Suppose that $\omega_b=0$ for some integer $b$, we have for all $i: 0\leq i< p-1$
$$
W^{(1)}_b(\theta^{2^i})=W^{(1)}_b(\theta)^{2^i}=0
$$
and
$$
1+W^{(2)}_b(\theta^{2^i})=(1+W^{(2)}_b(\theta))^{2^i}=(1+\omega_{b+f/2})^{2^i}=(\omega_b)^{2^i}=0,
$$
which tell us that both $W^{(1)}_b(X)$ and $1+W^{(2)}_b(X)$ have $p-1$ many solutions, since  2 is a primitive root modulo $p$. This contradicts to that one of $\deg(W^{(1)}_b(X))$ and $\deg(1+W^{(2)}_b(X))$ is smaller than $p-1$.

A similar proof can also lead to a contradiction if $\omega_b=1$. We complete the proof. \qed\\

For any non-zero polynomial $h(X)\in \mathbb{F}_2[X]$, the \emph{weight} of $h(X)$ is referred to as the number of non-zero coefficients of $h(X)$.

\begin{lemma}\label{weight-h-add}
Suppose that 2 is a primitive root modulo $p$.  Let $\theta\in \overline{\mathbb{F}}_2$ be a primitive $p$-th root of unity
 and $\omega_b=\sum\limits_{n\in \mathcal{Q}_b}\theta^n$ for any integer $b$. For any non-constant polynomial $h(X)\in \mathbb{F}_2[X]$
with $h(\theta)=\omega_b$, we have $wt(h(X))\geq (p-1)/2$.
 \end{lemma}
Proof.  By Lemma \ref{www-add} we see that  $\omega\in \overline{\mathbb{F}}_2\setminus\mathbb{F}_2$.
First, we can choose $h(X)\in \mathbb{F}_2[X]$ such that 
$h(X)=W^{(1)}_b(X)=\sum\limits_{n\in \mathcal{Q}_b}X^n$, then we have $h(\theta)=W^{(1)}_b(\theta)=\omega_b$.
In this case $wt(h(X))= (p-1)/2$.

Second, suppose that there is an $h_0(X)\in \mathbb{F}_2[X]$ such that $wt(h_0(X))< (p-1)/2$ and $h_0(\theta)=\omega_b$. Let $\overline{h}_0(X)\equiv h_0(X) \pmod {X^p-1}$ with $\deg(\overline{h}_0)<p$ and let
$H_0(X)=\overline{h}_0(X)+W^{(1)}_b(X)$, the degree of which is $<p$. Clearly $H_0(X)$ is non-zero since $\overline{h}_0(X)\neq W^{(1)}_b(X)$, since the weight of $W^{(1)}_b(X)$
is $(p-1)/2$.
Then we derive that $H_0(\theta)=0$ and $H_0(\theta^{2^i})=0$ for
$1\le i<p-1$.  Since  2 is a primitive root modulo $p$, we see that
$$
(1+X+X^2+\ldots+X^{p-1})|H_0(X),
$$
i.e., due to $\deg(H_0(X))<p$, we have $H_0(X)= 1+X+X^2+\ldots+X^{p-1}$, from which we get $\overline{h}_0(X)=1+W^{(2)}_b(X)$ and $wt(\overline{h}_0(X))= (p+1)/2$.
Therefore, $wt(h_0)\geq wt(\overline{h}_0)= (p+1)/2$, a contradiction. \qed\\

Now we turn to prove our main result.

\section{Proof of the main theorem}\label{proofmain}

\textbf{(Proof of Theorem \ref{klc-2-primitive}).} From the construction (\ref{def-binary}), we see that the weight of $(s_n)$ is $(p^2-1)/2+1$, i.e., there are $(p^2-1)/2+1$ many 1's in one period. Changing
all terms of 0's of $(s_n)$ will lead to the constant 1-sequence, whose linear complexity is 1. And changing
all terms of 1's will lead to the constant 0-sequence. So we always assume that $k<(p^2-1)/2$.

The generating polynomial of $(s_n)$ is of the form
\begin{equation}\label{gene-poly}
S(X)=1+\sum\limits_{i=0}^{pf/2-1} d_{i+b\pmod {pf}}^{(p^2,f)}(X)+ \sum\limits_{i=0}^{f/2-1} d_{i+b\pmod {f}}^{(p,f)}(X^p) \in \F_2[X],
\end{equation}
where $d_{l}^{(p^j,f)}(X)$ is defined in Eq.(\ref{d-poly}).
We first recall the linear complexity of $(s_n)$ that $LC^{\F_2}_0((s_n))=p^2$ for $f=2^r$ ($r\geq 1$) from \cite[Thm.1]{XZLH} and for even $f$ from
\cite[Thm.8]{ELZH}.

Then we suppose $k>0$.
Let
\begin{equation}\label{Hk}
S_k(X)=S(X)+e(X)\in \F_2[X]
\end{equation}
be the generating polynomial of the sequence obtained from $(s_n)$ by changing exactly $k$ terms of $(s_n)$ per period,
where $e(X)$ is the corresponding \emph{error polynomial} with $k$ many  monomials. We note that $S_k(X)$ is a nonzero polynomial due to $k<(p^2-1)/2$. We will consider the common roots of $S_k(X)$ and $X^{p^2}-1$,
i.e.,  the roots of the form $\beta^{n}  ~ (n\in \Z_{p^2})$ for $S_k(X)$, where $\beta \in \overline{\mathbb{F}}_{2}$ is a  primitive $p^2$-th root of unity. The number of the common roots will help us to derive the values of $k$-error linear
complexity of $(s_n)$  by  Eq.(\ref{licom}).\\

Case I: $k <(p^2-p)/2$.

On the one hand, we assume that $S_k(\beta^{n_0})=0$ for some $n_0\in \Z_{p^2}^*=\{1\leq n<p^2 | \gcd(p,n)=1\}$. Since $2$ is a primitive root modulo $p^2$, for each $n\in\Z_{p^2}^*$, there exists a $0\le j_n<(p-1)p$ such that $n\equiv n_02^{j_n} \pmod {p^2}$. Then we have
$$
S_k(\beta^{n})=S_k(\beta^{n_0 2^{j_n}})=S_k(\beta^{n_0})^{2^{j_n}}=0,
$$
that is, all ($p^2-p$) many elements $\beta^n$ with $n\in\Z_{p^2}^*$ are roots of $S_k(X)$. Hence we have
$$
\Phi(X)|S_k(X) ~~ \mathrm{in}~ \overline{\F}_2[X],
$$
where
$$
\Phi(X)=1+X^p+X^{2p}+\ldots+X^{(p-1)p}\in \F_2[X],
$$
the roots of which are exactly $\beta^n$ for $n\in\Z_{p^2}^*$.  Let
\begin{equation}\label{pi}
S_k(X)\equiv \Phi(X)\pi(X) \pmod {X^{p^2}-1}.
\end{equation}
Since $\deg(S_k(X))=\deg(\Phi(X))+\deg(\pi(X))< p^2$, we see that  $\pi(X)$ should be one of the following:
$$
\pi(X)=1;
\pi(X)=X^{v_1}+X^{v_2}+\ldots+X^{v_{t}};
\pi(X)=1+X^{v_1}+X^{v_2}+\ldots+X^{v_{t}};
$$
where $1\le t< p$ and $1\le v_1<v_2<\ldots<v_{t}<p$.
 Then the exponent of each monomial in $\Phi(X)\pi(X)$ forms the set
$\{lp : 0\le l\le p-1\}$ or $\{v_j+lp : 1\le j\le t, 0\le l\le p-1\}$ or
$\{lp, v_j+lp : 1\le j\le t, 0\le l\le p-1\}$ for different $\pi(X)$ above.

(i). If $\pi(X)=1$, we see that by (\ref{gene-poly})-(\ref{pi})
$$
e(X)=\sum\limits_{j=0}^{pf/2-1} d_{b+j\pmod {pf}}^{(p^2,f)}(X)+ \sum\limits_{n\in \mathcal{N}_b}X^{np}
$$
which implies that $k=(p^2-1)/2$, where $\mathcal{N}_b$ is in Eq.(\ref{QN}).

(ii). If $\pi(X)=X^{v_1}+X^{v_2}+\ldots+X^{v_{t}}$, we let
$$
\mathcal{I}=\{v_1, v_2,\ldots, v_{t}\}, ~~ \mathcal{J}=\{1,2,\ldots,p-1\}\setminus \mathcal{I}
$$
and let $z=|\mathcal{I}\cap \mathcal{Q}_b|$. We have $|\mathcal{J}\cap \mathcal{Q}_b|=(p-1)/2-z$, $|\mathcal{I}\cap \mathcal{N}_b|=t-z$ and
$|\mathcal{J}\cap \mathcal{N}_b|=(p-1)/2-t+z$.
For $\mathcal{V}_v=\{v+\ell p : 0\le \ell <p\}$,
by Lemma \ref{card-CC-add} we  derive that
$$
\begin{array}{rl}
e(X) = & \sum\limits_{v\in \mathcal{I}\cap \mathcal{Q}_b}~~\sum\limits_{n\in \mathcal{V}_v\cap \mathcal{C}_0}X^{n} +
\sum\limits_{v\in \mathcal{J}\cap \mathcal{Q}_b}~~\sum\limits_{n\in \mathcal{V}_v\cap \mathcal{C}_1}X^{n}
 \\
& + \sum\limits_{v\in \mathcal{I}\cap \mathcal{N}_b}~~\sum\limits_{n\in \mathcal{V}_v\cap \mathcal{C}_0}X^{n} +
\sum\limits_{v\in \mathcal{J}\cap \mathcal{N}_b}~~\sum\limits_{n\in \mathcal{V}_v\cap \mathcal{C}_1}X^{n} +1+ \sum\limits_{n\in \mathcal{Q}_b} X^{np},
\end{array}
$$
which implies that
$$
\begin{array}{rl}
k=& z\cdot (p-1)/2+((p-1)/2-z)\cdot (p+1)/2 +(t-z)\cdot (p+1)/2\\
&+((p-1)/2-t+z)\cdot (p-1)/2 + 1+(p-1)/2\\
=&(p^2-1)/2+1+t-2z.
\end{array}
$$
We can check that $-(p-1)/2\leq t-2z\leq (p-1)/2$. Then $k\geq (p^2-p)/2+1$.

(iii). For $\pi(X)=1+X^{v_1}+X^{v_2}+\ldots+X^{v_{t}}$, we can get similarly
$$
\begin{array}{rl}
e(X) = & \sum\limits_{v\in \mathcal{I}\cap \mathcal{Q}_b}~~\sum\limits_{n\in \mathcal{V}_v\cap \mathcal{C}_0}X^{n} +
\sum\limits_{v\in \mathcal{J}\cap \mathcal{Q}_b}~~\sum\limits_{n\in \mathcal{V}_v\cap \mathcal{C}_1}X^{n}
 \\
& + \sum\limits_{v\in \mathcal{I}\cap \mathcal{N}_b}~~\sum\limits_{n\in \mathcal{V}_v\cap \mathcal{C}_0}X^{n} +
\sum\limits_{v\in \mathcal{J}\cap \mathcal{N}_b}~~\sum\limits_{n\in \mathcal{V}_v\cap \mathcal{C}_1}X^{n} + \sum\limits_{n\in \mathcal{N}_b} X^{np},
\end{array}
$$
and $k=(p^2-1)/2+t-2z$. As in (ii), we have $k\geq (p^2-p)/2$.

So putting everything together, if $k< (p^2-p)/2$, we always have
$S_k(\beta^{n})\neq 0$ for all $n\in \Z_{p^2}^*$.\\

 On the other hand, we note first that $p\equiv -3 \pmod 8$ or $p\equiv 3 \pmod 8$ when  $2$ is a primitive root modulo $p^2$.
 From Eq.(\ref{gene-poly}) we have by Lemmas \ref{Dl} and \ref{card-CC-add}
 $$
 S(X)\equiv 1+\frac{p-1}{2}\sum\limits_{n=1}^{p-1}X^{n}+ \sum\limits_{n\in \mathcal{Q}_b}X^n+ \frac{p-1}{2}  \pmod{X^p-1}.
 $$
So let $\theta=\beta^p$, we get for $0\leq i<p$
\begin{equation}\label{roots}
\begin{array}{rl}
S_k(\beta^{ip})= & S_k(\theta^i)= e(\theta^{i})+S(\theta^i)\\
=& e(\theta^{i})+\left\{
\begin{array}{ll}
(p^2+1)/2, & \mathrm{if}~ i=0 ,\\
1+\sum\limits_{n\in \mathcal{Q}_b}\theta^{ni}, & \mathrm{if}~ 1\leq i<p,
\end{array}
\right.\\
=& e(\theta^{i})+\left\{
\begin{array}{ll}
1, & \mathrm{if}~ i=0 ,\\
1+\sum\limits_{n\in \mathcal{Q}_{b+\ell}}\theta^{n}, & \mathrm{if}~ i\in D_\ell^{(p,f)}, 0\leq \ell<f,
\end{array}
\right.\\
=& e(\theta^{i})+\left\{
\begin{array}{ll}
1, & \mathrm{if}~ i=0 ,\\
1+\omega_{b+\ell}(=\omega_{b+\ell+f/2}), & \mathrm{if}~ i\in D_\ell^{(p,f)}, 0\leq \ell<f,
\end{array}
\right.
\end{array}
\end{equation}
where $\omega_b=\sum\limits_{n\in \mathcal{Q}}\theta^n$. 
We want to look for an $e(X)$ with the least $wt(e(X))$ such that
 $e(1)=1$ and $e(\theta^{i})=\omega_{b+\ell+f/2}$ for $i\in D_\ell^{(p,f)}$.
These help us to calculate the number of roots for $S_k(X)$ of the form $\beta^{ip}  ~ (0\le i<p)$.

First, we consider the case $1\leq k<(p-1)/2$. If  $e(X)$ is with $1\le wt(e(X))<(p-1)/2$, we have
$e(1)=wt(e(X))$ and $e(\theta^{i})\neq \omega_{b+\ell+f/2}$ for $1\le i<p$ by Lemma  \ref{weight-h-add}. So we can use any monomial $e(X)$ (i.e., $wt(e(X))=1$) to deduce
$S_k(\beta^{0})=0$ but $S_k(\beta^{ip})\neq 0$ for $1\le i<p$.
So by Eq.(\ref{licom}) we derive
$$
LC^{\F_2}_{(p-3)/2}((s_n))=LC^{\F_2}_1((s_n))=p^2-1.
$$

Second, we consider the case $k=(p-1)/2$. Let $e(X)$ satisfy $wt(e(X))=(p-1)/2$ and $e(X)\equiv \sum\limits_{n\in \mathcal{N}_b}X^n \pmod{X^p-1}$.
Then we have  $e(1)=(p-1)/2$ and 
$$
 e(\theta^{i})=
\sum\limits_{n\in \mathcal{N}_b}\theta^{ni}=\sum\limits_{n\in \mathcal{N}_{b+\ell}}\theta^n=\omega_{b+\ell+f/2}  ~~~\mathrm{for}~~~
 1\le i<p,
$$
if  $i\in D_\ell^{(p,f)}$ for $0\leq \ell<f$. For such $e(X)$, it indicates that  
 $S_k(\beta^{0})=(p+1)/2$ and $S_k(\beta^{ip})= 0$ for $1\le i<p$. Hence
$$
LC^{\F_2}_{(p-1)/2}((s_n))=\left\{
\begin{array}{ll}
p^2-p+1, & \mathrm{if}~ p\equiv -3 \pmod 8,\\
p^2-p, & \mathrm{if}~ p\equiv 3 \pmod 8.
\end{array}
\right.
$$
Furtherly, for the case when $p\equiv 5 \pmod 8$, we choose an $e(X)$ satisfying $wt(e(X))=(p+1)/2$ and $e(X)\equiv 1+\sum\limits_{n\in \mathcal{Q}_b}X^n \pmod{X^p-1}$.
Then we derive $S_k(\beta^{ip})= 0$ for $0\le i<p$ and hence
$$
LC^{\F_2}_{(p+1)/2}((s_n))=p^2-p, ~~~\mathrm{if}~~~ p\equiv -3 \pmod 8.
$$\\

Case II: $k =(p^2-p)/2$.

Now we consider the case $k= (p^2-p)/2$. From (iii) above, only that $\mathcal{I}=\mathcal{Q}_b$ is useful for us, in this case $t=z=(p-1)/2$ and
$$
e(X) =\sum\limits_{v\in \mathcal{Q}_b}~~\sum\limits_{n\in \mathcal{V}_v\cap \mathcal{C}_0}X^{n} +
\sum\limits_{v\in \mathcal{N}_b}~~\sum\limits_{n\in \mathcal{V}_v\cap \mathcal{C}_1}X^{n} + \sum\limits_{n\in \mathcal{N}_b} X^{np},
$$
which can guarantee $S_k(\beta^{n})= 0$ for all $n\in \Z_{p^2}^*$. We also check that by Lemma \ref{card-CC-add}
$$
\begin{array}{rl}
e(\beta^{ip})= e(\theta^i)= & \frac{p-1}{2} \sum\limits_{n\in \mathcal{Q}_b} \theta^{ni}+
\frac{p-1}{2} \sum\limits_{n\in \mathcal{N}_b} \theta^{ni}+\frac{p-1}{2} \\
= & \frac{p-1}{2} \left( \sum\limits_{n\in \mathcal{Q}_b} \theta^{ni}+
 \sum\limits_{n\in \mathcal{N}_b} \theta^{ni}\right)+\frac{p-1}{2} \\
 = & \frac{p-1}{2} +\frac{p-1}{2}=p-1=0
\end{array}
$$
for $1\le i<p$ and $e(\beta^0)=e(1)=(p-1)/2$. Then from Eq.(\ref{roots}), we get
$S_k(\beta^{0})=(p+1)/2$ and $S_k(\beta^{ip})\neq 0$ for  $1\le i<p$. So we have
$$
LC^{\F_2}_{(p^2-p)/2}((s_n))=p-\delta,
$$
where $\delta=1$ if  $p\equiv 3 \pmod 8$ and $\delta=0$ if $p\equiv -3 \pmod 8$.\\

Case III: $k >(p^2-p)/2$.

If   $k=(p^2-1)/2$, after changing the $k$ many 0's in $(s_n)$,  we  get the 1-sequence, whose linear complexity is 1.
And if   $k>(p^2-1)/2$, we can get the 0-sequence whose linear complexity is 0.
We complete the proof.   \qed\\

We remark that, it seems difficult for us to determine  the values of $LC^{\F_2}_{k}((s_n))$ for $(p^2-p)/2<k<(p^2-1)/2$ here,
but it is at most $p$ (or $p-1$).

\section{A lower bound}

We have the following lower bound on the $k$-error linear complexity when $2$ is not a primitive root modulo $p^2$.

\begin{theorem}\label{lowerbound}
Let $(s_n)$ be the binary sequence of period $p^{2}$ defined in Eq.(\ref{def-binary}) with even $f$ and any $b$ for defining $\mathcal{C}_0$ and $\mathcal{C}_1$. 
If $2^{p-1} \not \equiv 1 \pmod{p^2}$, then
the $k$-error linear complexity  of $(s_n)$  satisfies
$$
LC^{\F_2}_k((s_n))\geq \lambda p ~~~~ \mathrm{for}~~~ 0\leq k<(p^2-p)/2,
$$
where $1<\lambda< p$ is the order of $2$ modulo $p$.
\end{theorem}

Proof. First we show the order of $2$ modulo $p^2$ is $\lambda p$. Under the assumption on  $2^{p-1} \not \equiv 1 \pmod{p^2}$, we see that  the order of $2$ modulo $p^2$ is
of the form $mp$ for some $1\le m\le p-1$ and $m|(p-1)$. Then $\lambda | m$ from $1\equiv 2^{mp} \equiv 2^{m} \pmod{p}$ since $2^{mp} \equiv 1 \pmod{p^2}$,
and $m | \lambda$ from $ 2^{\lambda p} \equiv 1 \pmod{p^2}$ since otherwise we write for some $1\le \epsilon<p$
$$
2^{\lambda p} \equiv 1 + \epsilon p \pmod{p^2},
$$
from which we derive
$$
1\equiv (2^{\lambda p})^{m/\lambda} \equiv (1 + \epsilon p)^{m/\lambda}\equiv 1 + \frac{\epsilon  m p}{\lambda}  \pmod{p^2}.
$$
However $\frac{\epsilon  m}{\lambda}\not\equiv 0   \pmod{p}$, a contradiction.

Second, the fact $ k<(p^2-p)/2$ implies that there do exist an $n_0\in \Z_{p^2}^*$ such that
 $S_k(\beta^{n_0})\neq 0$, where $S_k(X)$, as before, is the generating polynomial of the sequence obtained from $(s_n)$ by changing exactly $k$ terms per period.
Since otherwise, $ k\geq (p^2-p)/2$ according to Case I in the proof of Theorem \ref{klc-2-primitive}.

Thus there are at least $\lambda p$ many $n\in \{n_02^j \bmod {p^2} : 0\le j<\lambda p\}$ such that $S_k(\beta^{n})\neq 0$. Then the result follows.\qed\\

Theorem \ref{lowerbound} covers almost all primes. As far as we know, the primes satisfying $2^{p-1}   \equiv 1 \pmod{p^2}$ are very rare.
It was shown  that there are only two such primes\footnote{A prime $p$ satisfying $2^{p-1}   \equiv 1 \pmod{p^2}$ is called a Wieferich prime.}, 1093 and 3511,
up to $6 \times 10^{17}$ \cite{AS}.\\

Finally, we draw a conclusion that we have determined the values of the $k$-error linear complexity of a new generalized cyclotomic binary sequence of period $p^2$
discussed recently in the journal Designs, Codes and Cryptography.
Results indicate that such sequences  have large linear complexity and
 the linear complexity does not significantly decrease by changing a few
terms.

\section*{Acknowledgements}

The work was partially supported by the National Natural Science
Foundation of China under grant No.~61772292, by the Provincial Natural Science
Foundation of Fujian under grant No.~2018J01425 and by 2016 Development Program for Distinguished Young Scientific Research Talent of Universities in Fujian Province.

C. X. Xu was also partially supported by  the National Key R\&D Program of China under grant No.~2017YFB0802000.

P. H. Ke was also partially supported by the National Natural Science
Foundation of China under grant No.~61772476.


\begin{thebibliography}{99}


\bibitem{AS} A. Akbary, S. Siavashi. The largest known Wieferich numbers. Integers, 18-\#A3 (2018)  1-6.


\bibitem{CZHTY13} H. Cai, X. Zeng, T. Helleseth, X. Tang, Y. Yang.
 A new construction of zero-difference balanced functions and its applications. IEEE Trans. Inf. Theory 59(8) (2013) 5008-5015.



\bibitem{Chen}  Z. X. Chen. Trace representation and linear complexity of binary sequences derived from Fermat quotients.  Sci. China  Inf. Sci., 57 (2014) 11:2109(10)



\bibitem{CD}  Z. X. Chen  and  X. N. Du. On the linear complexity of binary threshold sequences derived from
Fermat quotients. Des. Codes Cryptogr., 67 (2013) 317-323.



\bibitem{CG}Z. X. Chen  and  D. G\'{o}mez-P\'{e}rez. Linear complexity of
binary sequences derived from polynomial quotients. Sequences and Their Applications-SETA 2012, 181-189, Lecture Notes in Comput. Sci., 7280, Springer, Berlin, 2012.


\bibitem{CNW}  Z. X. Chen, Z. H. Niu and C. H. Wu. On the $k$-error linear complexity of binary
sequences derived from polynomial quotients. Sci. China  Inf. Sci.,  58  (2015) 09:2107(15)


 \bibitem{COW}  Z. X. Chen, A. Ostafe and  A. Winterhof. Structure of
 pseudorandom numbers derived from Fermat quotients. Arithmetic of Finite Fields-WAIFI 2010, 73-85, Lecture Notes in Comput. Sci., 6087, Springer, Berlin, 2010.






\bibitem{CDR}
 T. W. Cusick, C. Ding and A. Renvall. Stream Ciphers and Number Theory. Gulf Professional Publishing, 2004.


\bibitem{DH98} C. Ding, T. Helleseth.  New generalized cyclotomy and its applications. Finite Fields
Appl. 4 (1998) 140-166.



\bibitem{DH99} C. Ding, T. Helleseth.  Generalized cyclotomy codes of length $p_1^{m_1}p_2^{m_2}\cdots p_t^{m_t}$. IEEE
Trans. Inf. Theory 45(2) (1999) 467-474.

\bibitem{DXS}
 C. S. Ding, G. Z. Xiao,  W. J. Shan.  The stability theory of stream
ciphers. Lecture Notes in Computer Science, 561. Springer-Verlag,
Berlin, 1991.

 \bibitem{ELZH}
V. Edemskiy, C. L. Li, X. Y. Zeng, T. Helleseth.
The linear complexity of new binary cyclotomic sequences of period $p^n$. Preprint, 2018 (personal communication)



\bibitem{GW} D. G\'{o}mez-P\'{e}rez and A. Winterhof. Multiplicative character sums of
Fermat quotients and pseudorandom sequences, Period. Math. Hungar.,
64 (2012) 161-168.


\bibitem{OS} A. Ostafe  and I. E. Shparlinski. Pseudorandomness and dynamics of
Fermat quotients. SIAM J. Discr. Math., 25 (2011)  50-71.



\bibitem{SM} M. Stamp and C. F. Martin. An algorithm for the $k$-error linear complexity of
 binary sequences with period $2^n$. IEEE Trans. Inform. Theory, 39 (1993)  1398-1401.

\bibitem{W62}A. L. Whiteman. A family of difference sets. Illionis J. Math., 6(1) (1962) 107-121.


\bibitem{XZLH} Z. B. Xiao, X. Y. Zeng, C. L. Li, T. Helleseth.
New generalized cyclotomic binary sequences of period $p^2$. Designs, Codes and Cryptography, https://doi.org/10.1007/s10623-017-0408-7.(2017)


\bibitem{XCX16} S. Xu, X. Cao, G. Xu. Optimal frequency-hopping sequence sets based on cyclotomy. Int. J. Found. Comput. Sci., 27(4) (2016) 443-462.

\bibitem{ZCTY13} X. Zeng, H. Cai, X. Tang, Y. Yang. Optimal frequency hopping sequences of odd
length. IEEE Trans. Inf. Theory, 59(5) (2013) 3237-3248.

\end{thebibliography}
\end{document}